# Raman scattering study of lattice and magnetic excitations in CrAs

K. Sen[1,*], Y. Yao[1], R. Heid[1], A. Omoumi[1], F. Hardy[1], K. Willa[1], M. Merz[1], A. A. Haghighirad[1], and M. Le Tacon[1,†]

[1]Institute for Solid State Physics, Karlsruhe Institute of Technology, D-76344, Eggenstein-Leopoldshafen, Germany

We report on the lattice and spin dynamics in the monopnictide CrAs using polarized Raman scattering. This system exhibits a first-order magneto-structural phase transition at $T_N \sim$ 265 K, below which the magnetic moments of Cr form an incommensurate double helical magnetic structure while the unit cell volume abruptly expands. Across the transition, the frequencies of the fully symmetric $A_g$ phonons strongly renormalize, which along with the results of first-principle calculations suggest the presence of a sizeable coupling of the phonons to the magnetic degrees of freedom in this compound. In addition, we observe two broad modes around 350 and 1700 cm$^{-1}$ in the magnetic phase, which we associate with two-magnon Raman scattering. This work provides one of the few examples of magnetic Raman scattering from a non-collinear magnet.

## I. Introduction

Transition-metal pnictides have been widely studied in the recent years, in particular following the discovery of high-temperature superconductivity in the Fe-based compounds [1–4]. This has also revived the interest for related compounds including monopnictides such as MnAs, which has long been known as a magneto-caloric material [5]. This paper focusses on another representative of this family of compounds, CrAs, a semimetal which undergoes a first-order phase transition from a paramagnetic (PM) state to a non-collinear magnetic state below $T_N \approx 265$ K at ambient pressure [6–9]. At $T_N$, it exhibits a giant structural anomaly which involves a large expansion of the crystallographic *b*-axis by ~3.9 % and the unit cell volume by ~2.2 %, while the space group of *Pnma* remains unaltered [10].

The itinerant character and the complex antiferromagnetic (AF) magnetic structure of CrAs have attracted some attention in the early 70s [6,7,9,11,12]. In particular, the magnetic structure has been investigated in great details, and it revealed a large magnetic moment of Cr (~1.73 $\mu_B$ at 1.5 K) lying in the *ab* crystallographic plane and forming a double helical structure with the propagating wave vector (0, 0, 0.356) in reciprocal lattice unit [8]. Studies on single crystals have overall been limited by the fact that pristine compound tends to break upon thermal cycling, which is due to the large magnetostriction effect across $T_N$. Yet, the interest for this material has recently been revived for several reasons, the first of them being the discovery of superconductivity (SC) under pressure in this system. When the helical magnetic order is suppressed under hydrostatic pressure, bulk SC emerges and displays a dome-like pressure dependence with a maximum SC transition temperature of $T_c \approx 2$ K [13–15], making this system the first known 'Cr-based' superconductor.

Whilst the dome-like pressure dependence of $T_c$ is strongly reminiscent of unconventional superconductors, CrAs exhibits several other features that are typically encountered in correlated materials. For instance, in the paramagnetic (PM) phase above $T_N$, the magnetic susceptibility increases linearly with temperature up to 700 K [16], similar to what is generically found in iron-based superconductors, and often regarded as an evidence for finite AF fluctuations in the PM phase [14].

Below $T_N$, the electrical resistance drops abruptly and follows a Fermi-liquid-like $T^2$ behavior [14,17,18], suggesting a substantial modification of the Fermi surface across the magneto-structural transition. Strong deviations from this Fermi-liquid behavior are observed as the long-range magnetic order entirely vanishes above a critical pressure of $p_c \approx 10$ kbar, and are accompanied with a sizeable enhancement of the electronic effective mass close to $p_c$ (inferred from specific heat measurements). This has been interpreted as a signature of quantum criticality, which in turn suggests an unconventional superconducting pairing mechanism [17]. Furthermore, it has been argued that the AF spin fluctuations that are present at ambient pressure are critically enhanced at $\approx p_c$ [14]. Around this pressure, quasi-linear and non-saturating magnetoresistance have been reported and are associated with the singular electronic band structure of this material [19]. Above $p_c$, nuclear quadrupole resonance experiments revealed the presence of substantial magnetic fluctuations [20], whereas the electrical resistivity depends quasi-linearly on the temperature [17]. Already at ambient pressure, the very large magnetostriction effect at $T_N$ indicates a strong coupling of the lattice to the underlying static magnetic order. To the best of our knowledge, however, no experimental study of the lattice dynamics of CrAs nor of its coupling to magnetism has been reported.

Here, we address this topic and present a comprehensive Raman scattering study of lattice dynamics and magnetic excitations in CrAs as a function of temperature at ambient pressure. Raman spectroscopy is an ideal tool to simultaneously probe zone-center lattice (phonons), and magnetic (single and bi-magnons) excitations in correlated systems [21]. We reveal anomalously large renormalization (both in frequency and linewidth) of the phonons of $A_g$ symmetry across the magneto-structural transition, as well as some excitations below $T_N$ that we associate with two-magnon Raman scattering. Detailed comparison with first-principle calculations indicates the presence of a large spin-phonon coupling in CrAs.

The paper is organized as follows. Section II comprises details on the crystal growth, structural and magnetic characterization, Raman scattering measurements and the corresponding data analyses. In section III, we present the details of the first-principle calculations. Experimental results and discussion are presented in sections IV and V, respectively.

## II. Experiment

### A. Crystal growth and structural characterization

Crystals of CrAs were grown from Sn-flux. First, polycrystalline CrAs was prepared from high-purity elements chromium (99.99 %, Alfa Aesar) and arsenic (99.999 %, ChemPUR) by weighing the stoichiometric mixture of the elements, which was subsequently evacuated and sealed in a quartz ampoule. The latter was heated at 700 °C for 20 h and then cooled down to room-temperature (RT) at a rate of 50 °C/h. The product was ground in argon atmosphere in a glove box and the phase purity was checked by powder x-ray diffraction. CrAs and pieces of Sn with molar ratio of CrAs:Sn ≈ 1:10 were added to an alumina crucible and placed in a quartz ampoule which was evacuated at $10^{-5}$ mbar and sealed. The quartz ampoule was heated up to 650 °C and kept there for 10 h, then heated up to 900 °C with a dwell time of 5 h and slowly cooled down to 685 °C in 7 days. At this temperature, the excess of Sn was centrifuged and the remaining Sn-flux was washed off the CrAs single crystals using a 30 % HNO$_3$ solution. Large crystals with shiny facets and dimensions up to 5 mm × 0.5 mm × 0.7 mm (see Fig. 1(a)) were obtained. The single crystals have habitus that look similar to a hexagonal system. This can be explained by the fact that CrAs undergoes a second-order phase transition below 1173 K from a NiAs-type (hexagonal) structure to a MnP-type (orthorhombic) structure [12,22].

Temperature-dependent x-ray diffraction data on CrAs single crystals were collected between 85 K and room temperature on a Stoe imaging plate diffraction system (IPDS-2T) using Mo $K_\alpha$ radiation. All accessible symmetry-equivalent reflections (≈ 5000) were measured up to a maximum angle of $2\theta = 65°$.

The data were corrected for Lorentz, polarization, extinction, and absorption effects. Using SHELXL [23] and JANA2006 [24] around 160 averaged symmetry-independent reflections (I > 2σ) have been included for the respective refinements in space group *Pnma*. The refinement converged quite well and shows excellent weighted reliability factors ($wR_2$) which are for all temperatures around 4 %. The temperature-dependent lattice parameters, bond angles and bond distances were derived from the atomic positions of Cr and As. They are shown in Fig. 1(d), Fig. 6(b), and Figs. 6(c)-(d), respectively. In addition, we listed out structural parameters at room-temperature and 95 K in Appendix A.

### B. Magnetization measurements

The magnetic susceptibility of CrAs was obtained by gluing a single crystal with a mass of ≈12 mg to a polyether ether ketone substrate and mounting it into a Quantum Design-Physical Properties Measurement System. Measurements were performed with the standard Vibrating Sample Magnetometer option at a field of 1 T along the crystallographic *a*-axis. The measured susceptibility, plotted in Fig. 1(c), shows a sharp jump at the magnetic transition with a width of <1.5 K. We observe a pronounced hysteresis, as expected for a first-order phase transition, with transition temperatures of 265 K upon cooling and 272 K on warming.

### C. Raman scattering

Raman scattering experiments were performed in backscattering geometry using a Jobin-Yvon LabRAM HR Evolution spectrometer. A He-Ne laser (λ=632.8 nm) was focused onto the facets of the single crystals (see Fig. 1(a)) through a 50× microscope objective. We used a laser power of ≈ 1 mW and a laser spot size of ≈ 5 μm in diameter to limit laser-induced heating. The scattered photons are dispersed by a grating with 1800 (respectively 600) grooves/mm and recorded with a Peltier-cooled CCD detector. The total energy resolution of the spectrometer is 0.6 (resp. 1.8) cm$^{-1}$ which was determined from the direct measurements of resolution limited Ne emission-lines. Temperature dependent Raman spectra were acquired by cooling the sample in a He-flow cryostat. The Raman susceptibility, $\chi''$, was obtained by correcting the recorded Raman spectra from the Bose factor, $1 + n(\omega, T)$.

The high-resolution mode was used to analyze the phonons, whereas the magnetic excitation spectra were recorded using the low-resolution mode, which maximizes the throughput of the Raman spectrometer. Phonon frequencies and linewidths were obtained by fitting the peaks using Voigt profiles (intrinsic Lorentzian lineshapes convoluted with the Gaussian experimental resolution).

From a group theoretical analysis of the orthorhombic structure of CrAs (space group: *Pnma*; point group: $D_{2h}$ - both Cr and As occupy 4c Wyckoff positions) [11,25,26], twenty-one optical phonon modes at the zone center are expected. Among them twelve are Raman active with the $A_g$ (4 modes), $B_{1g}$ (2), $B_{2g}$ (4) and $B_{3g}$ (2) symmetries.

The weak intensity of the $B_{1g}$ and $B_{2g}$ modes in crossed polarization (see in Appendix B) did not allow us to carry out their systematic temperature dependence. For this reason, in this paper, we have restricted ourselves to the study of the fully symmetric $A_g$ modes, which were probed by keeping the incoming and scattered photon polarizations parallel to the crystallographic *a*-axis of the sample ('XX' configuration in Porto's notation), irrespective of the facet of the crystal on which they are measured (see in Appendix C).

### III. Lattice dynamics calculation with density functional theory

Density-functional calculations were carried out for CrAs using the mixed-basis pseudopotential method [27]. We used norm-conserving pseudopotentials with nonlinear core-corrections, and included the semi-core states Cr-3s, Cr-3p, and As-4s in the valence space. The mixed-basis approach describes valence states with a combination of plane waves and local functions. The latter allow an efficient description of more localized components of the valence states. For CrAs, we used plane waves up to a kinetic energy of 24 Ry (326.5 eV), augmented by local functions of s, p, d type at the Cr sites. For the exchange-correlation functional, we employed the Perdew-Burke-Ernzerhof (PBE) parameterization of general-gradient approximation (GGA) [28], and Brillouin zone integrations were performed with a 8×12×8 orthorhombic k-point grid in conjunction with a Gaussian smearing of 0.1 eV.

Zone-centered phonons were calculated using the linear response or density-functional perturbation theory implemented in the mixed-basis scheme [29]. Internal structural parameters were relaxed prior to the phonon calculations. To analyze the effect of purely structural changes at the phase transition on the Raman modes, we performed calculations for two sets of experimental structural parameters corresponding to room-temperature and 95 K (see in Appendix A) respectively, after internal relaxation of the atomic positions. To study the effect of magnetism on the structure and on the lattice dynamics we performed additional spin-polarized calculations with both collinear ferromagnetic (FM) and antiferromagnetic (AF) structures. The AF order involves the antiparallel alignment of the neighboring Cr-spins within a unit cell both along in-plane and out-of-plane directions.

The obtained atomic displacement patterns of these individual modes are depicted in the inset of Fig. 2. The calculated eigenfrequencies of the $A_g$ Raman modes are shown in Figs. 5(a) and (b).

### IV. Experimental Results

#### A. Raman-active phonons

In Fig. 2, we show the typical Raman response of CrAs at two temperatures above and below the magneto-structural transition. At 310 K, all four expected $A_g$ modes were detected, and will be referred to $A_g$-1 (118.9 cm$^{-1}$), $A_g$-2 (178.8 cm$^{-1}$), $A_g$-3 (227.6 cm$^{-1}$) and $A_g$-4 (265.0 cm$^{-1}$), respectively. The Raman spectra are dominated by the $A_g$-1 mode which is almost an order of magnitude more intense than any of the other three modes. All the modes could be fitted with Voigt profiles, and no Fano asymmetry was detected.

As already evident from the data displayed in Fig. 2, the four Raman modes are strongly renormalized across $T_N$. In Figs. 3(a)-(h), we present a more detailed temperature dependence of the frequency and linewidth (full-width-at-half-maximum (FWHM)) of all the $A_g$ modes.

The temperature dependence of the phonon frequencies and linewidths in crystals is generally governed by anharmonic effects. Phonons are expected to harden continuously as the temperature decreases and the lattice contracts; whereas, their linewidths, inversely proportional to the modes´ lifetimes, decreases with decreasing phonon population and phonon-phonon scattering. We further fitted (solid lines in Fig. 3) the regular temperature dependence of the phonon frequencies and linewidths using the symmetric anharmonic phonon decay model (Klemens model), in which an optical phonon decays into two acoustic phonons of opposite momentums [30,31].

This trend for the frequency is qualitatively followed for all the modes above and below $T_N$. Across $T_N$, however, a sudden renormalization of the phonon frequencies is observed, as expected from the first-order nature of the magneto-structural transition. While the $A_g$-1, $A_g$-3 and $A_g$-4 modes significantly soften below the transition, the $A_g$-2 mode exhibits a small hardening $(\Delta\omega/\omega = (\omega_{T<T_N} - \omega_{T>T_N})/\omega_{T>T_N} = 1.3\,\%)$. We note here that the softening of the $A_g$-1 mode is particularly strong, with a relative frequency change of $\Delta\omega/\omega = -11.6\%$ across $T_N$. This exceeds the expectations from Grüneisen´s law which relates the phonon frequencies and the unit cell volume by $\Delta\omega_i/\omega_i = -\gamma_i \Delta V/V$. For the $A_g$-1 mode, we get the Grüneisen parameter as $\gamma_{A_g-1} \approx 5.3$, a value which is significantly larger than the typical values (2-3) for Grüneisen parameters [32].

It is also interesting to compare the rate of change of the phonon frequency above and below $T_N$. To this aim, we added dotted lines to Figs. 3(a)-(d) at $< T_N$, representing the slopes d$\omega$/dT measured between room temperature and $T_N$. Except for $A_g$-2, the rate of hardening of all phonons strongly slows down in the magnetic phase. Interestingly, whilst the hardening of 2-4 cm$^{-1}$ for all the modes between $T_N$ and base temperature is rather normal, the changes of the same amplitude within only 30 K between room-temperature and $T_N$ appear remarkable. This anomalously large temperature dependence indicates an unusually large lattice anharmonicity in the PM phase.

As far as the linewidths are concerned (Figs. 3(e)-(h)), the very small residual width of the phonons (0.5 cm$^{-1}$ FWHM for $A_g$-2 mode and 2-4 cm$^{-1}$ for others, including the modes of $B_{1g}$ and $B_{2g}$ symmetries) at base temperature confirms the high quality of the investigated single crystals, which is in accordance with our XRD results. In addition, the lack of Fano asymmetry in the lineshape of these modes indicates weak coupling of these phonons to any electronic continuum in CrAs.

Except for the $A_g$-4 mode, which has an extremely weak intensity to reliably determine its linewidth, the temperature dependences of the linewidths of the other phonons are found to be remarkable. The $A_g$-2 and $A_g$-3 modes generally narrow down upon cooling as expected, but their FWHMs abruptly drop by more than 50 % across $T_N$. To our surprise, the lineshape of $A_g$-1 does not seem to be affected by the magneto-structural transition. Moreover, slightly below $T_N$, it starts to anomalously broaden upon cooling, a trend that we observed down to $T \approx 125$ K. It, however, rapidly narrows down to the base temperature upon further cooling.

Before we discuss the origin of the anomalous phonon behavior of $A_g$-1 in section V, we briefly mention another set of Raman active excitations that rises below $T_N$.

### B. Evidence of magnetic Raman scattering

As seen already in the overview presented in Fig. 2, two distinct broad features at $\approx 350$ and $1700$ cm$^{-1}$ that were not present at 310 K are visible at 10 K. In Fig. 4(a), we present a detailed temperature dependence of the lower energy mode, as well as, in Fig. 4(b), the temperature dependent part of the Raman response $\Delta\chi''(T) = \chi''(T) - \chi''(310\text{ K})$ in the same temperature range after subtraction of the phonon contributions. This clearly indicates that the 350 cm$^{-1}$ feature appears just below $T_N$, narrows down and rapidly grows in amplitude as the temperature is further reduced.

To be more quantitative, we show in Fig. 4(c) the spectral weight of $\Delta\chi''$ (integrated between 150 and 750 cm$^{-1}$) as a function of temperature. The integrated intensity appears to saturate below $\approx 80$ K. Apparently, such a temperature dependence is closely linked to the magnetic susceptibility ($\chi_v$) shown in Fig. 1(c). The solid line in Fig. 4(c) corresponds to $|\chi_v(T) - \chi_v(T_N)|$, which perfectly reproduces the temperature dependence of the 350 cm$^{-1}$ feature's intensity. This naturally suggests a magnetic origin of this excitation.

We finally note that at the lowest temperatures, when the 350 cm$^{-1}$ mode becomes sufficiently narrow, an additional excitation can be resolved around $\approx 230$ cm$^{-1}$, which is below the very sharp $A_g$-3 mode. All these features are completely absent from the Raman response measured in crossed polarization (one along the $a$-axis, and the other in the $bc$-plane. See in Appendix B). This is found to be consistent with expectations from the analysis of the Fleury-Loudon magnetic light scattering Hamiltonian (see in Appendix D).

### V. Discussion

**Phonon renormalization across $T_N$.** As described in the section IV.A, the Raman active phonons of CrAs are strongly renormalized across the first-order magneto-structural transition. To gain more insight, we have estimated the expected changes in the phonon frequencies using first-principle calculations (section III) constrained to the lattice parameters, which were determined experimentally at room-temperature and at 95 K $\ll T_N$ (see in Appendix A). We first carried out the calculations without considering the long-range magnetic order of the system, which was taken into account in a second step.

In Fig. 5, we compare the measured and calculated phonon frequencies in CrAs at 310 and 95 K. At both temperatures, significant differences between the theoretical and experimental frequencies are observed. A good agreement ($< 5$ %) is found for the $A_g$-3 and $A_g$-4 modes, whereas sizeable deviations from the experimental values are found for the $A_g$-1 (~15 to 18 % off depending on the temperature) and $A_g$-2 (~8 to 10 % off) modes.

To understand the origin of this discrepancy, it is instructive to compare the experimentally determined bond lengths and angles between the atoms in the unit cell with the calculated ones. Figs. 6(c) and (d) reveal that at both temperatures most bond lengths agree within $\approx \pm 2$ % with the experimental values. However, a noticeable discrepancy concerns the Cr-Cr bond #B6 (marked in Fig. 6(a)), which is 5 % too short in the calculation. Similarly, it underestimates the Cr-As-Cr angle #A1 (denoted in Fig. 6(a)) by 3.16°. Furthermore, the analysis of the eigendisplacements of the $A_g$-1 and $A_g$-2 modes reveals that these are the phonons which give rise to the largest modulation of the Cr-As-Cr angle #A1, which directly relates to the dominant magnetic exchange path ($J_{c2}$ in Fig. 7(c)). This in turn suggests that the inclusion of magnetism in the calculation is necessary to accurately account for the lattice dynamics in CrAs.

Performing phonon calculations with explicit inclusion of the actual double helical magnetic structure is beyond the scope of the present paper and will be the subject of subsequent theoretical studies.

To simply evaluate the impact of magnetism on the structural parameters and on the phonon frequencies, we have imposed FM and AF orders, relax the atomic positions in the unit cell and calculate the frequencies of the Raman phonons. In agreement with a previous study [33], we find that the GGA functional significantly overestimates the magnetic moment in CrAs. We obtained 2.4 (~2.3) $\mu_B$/Cr above and below $T_N$ for FM (AF) case.

From the structural point of view, the inclusion of magnetism systematically improves the agreement between the calculated bond lengths and angles with the experimentally determined ones. As far as the bond lengths are concerned, the agreement is particularly good when considering AF order, while the FM order induces deviations to the experimental case opposite to that obtained for the non-magnetic (NM) case. Indeed, as seen in Figs. 6(c) and (d), bond lengths that tended to be underestimated in the NM calculations are now slightly over estimated and vice-versa. The Cr-As-Cr angles #A2 and #A4 are marginally affected by magnetism, opposite to #A1 and #A3 on which it has a much more dramatic effect. The FM calculation significantly overestimates #A1, which is slightly underestimated by the AF calculation. The opposite trend is seen for #A3, underestimated in the FM case, overestimated in the AF one. In either case, however, the agreement between FM/AF calculations and experiment is better than that of the NM calculation.

Consequently, the calculated phonon frequencies are generally found in better agreement with the experiment when including magnetism. FM has only little effect on $A_g$-4 while $A_g$-1, $A_g$-2 and $A_g$-3 are strongly renormalized (see in Figs. 5(a) and (b)). The former two are then in excellent agreement with the experiment (within $\approx \pm 3$ %). The trend is a bit different for the AF calculation that mostly affects the $A_g$-2 and $A_g$-4 modes, leaving $A_g$-1 essentially unmodified with respect to the NM case.

The detail of the interplay between the structural and magnetic degrees of freedom is certainly complex, but qualitatively we note that the inclusion of magnetism in the first principle calculation has the strongest impact on the structural degrees of freedom that modulate exchange paths $J_{c2}$ (AF) and $J_{c1}$ (FM), which in turn dramatically affect the phonon frequencies. This demonstrates a significant coupling between spin and lattice degrees of freedom, and interestingly, this conclusion is valid both below and above the magnetic transition. This is strongly reminiscent of the case of iron-based superconductors [34,35], in which the phonon frequencies and dispersion calculations agree best with the experiment when including magnetic order, even in the PM phase. We can finally expect that taking into consideration the actual double helical magnetic structure, which compromises between FM and AF interactions but it currently beyond our computational capabilities for lattice dynamic calculations, will yield the best agreement with the experiment.

We end this part of the discussion by noting that even in the non-magnetic calculation, the relative change $\Delta\omega/\omega = (\omega_{95K} - \omega_{310K})/\omega_{310K}$ of the phonon frequencies is already well captured (not shown). In other words, even if magnetism must be included to correctly reproduce the phonon frequencies, the large Grüneisen parameter of the $A_g$-1 mode does not have a magnetic origin. Furthermore, our investigations reveal that this mode actually corresponds to the branch that gets completely soft at the zone boundary (M-point) of the high-temperature hexagonal unit cell, when the phase transition to the orthorhombic phase takes place at 1173 K (this M-point is then backfolded to the zone-center). The amplitude of the structural orthorhombicity, defined as $c/b$ ($=\sqrt{3}$ in the hexagonal phase), increases down to $T_N$ and suddenly changes its sign below the transition. The low temperature structure is energetically very close to the hexagonal phase, which naturally affects the soft branch, to which the $A_g$-1 mode belongs. The situation is very close to that of MnAs, which however goes back to the hexagonal phase at low-temperature, and is also predicted to exhibit a large spin-phonon interaction [36].

**Impact of spin-phonon coupling on phonon lifetimes.** Having established the importance of magnetism to calculate phonon frequencies even above $T_N$, we now discuss its possible effects on the linewidths. As we have seen, the FWHM exhibits a singular behavior, with a marked jump across $T_N$ for modes $A_g$-2 and $A_g$-3. This behavior is strongly reminiscent of that of the magnetization, but also of that of the electrical resistivity $\rho(T)$ reported in [14,17]. Interestingly, we note that this empirical relationship between $\rho(T)$ and the lineshape of Raman active phonons can be seen in other magnetic metals, in particular in the parent compounds of Fe-based superconductors such as $BaFe_2As_2$ [37]. There, it has qualitatively been understood in terms of reduction of the electron-phonon interaction caused by the opening of a spin-density-wave gap in the electronic density of states. In this context, further investigation on changes in the electronic structure of CrAs across $T_N$, as well as subsequent studies on P-doped CrAs (which strongly impacts $\rho(T)$ [17]) using e.g. Angle Resolved Photoemission Spectroscopy would be required to better understand the origin of these linewidth anomalies).

The linewidth of the $A_g$-1 mode, on the other hand, exhibits a completely different temperature dependence; a cusp-like shape with a maximum at $\approx 125$ K (Fig. 3(e)), a temperature that does not correspond to any known anomaly in the structure or the thermodynamic properties of CrAs. This in turn suggests that such a remarkable temperature dependence of the linewidth of $A_g$-1 arises from a compromise between opposite effects below $T_N$. We already established that this mode is highly sensitive to the underlying magnetism. Therefore, spin-phonon coupling appears as a possible damping mechanism below $T_N$, which is then partially compensated at low temperatures by regular lattice anharmonic effects. Notably, in the past, several studies have highlighted the impact of spin-phonon coupling on phonon frequencies (see e.g. [38–40]), but its effect on the phonon linewidths has been barely discussed. Recent experimental studies on antiferromagnetic $Pr_{1-x}Ca_xMnO_3$ [41], $SmCrO_3$ and $GdCrO_3$ [42,43] reported similar phonon broadenings below $T_N$ and point towards a generic phenomenology for which a robust theoretical framework is missing to date.

**Magnetic Raman scattering.** The excitations around 350 and 1700 cm$^{-1}$ are only detected below $T_N$ and in the absence of broken translational symmetry, so that they are very likely to have magnetic origin. Single magnon Raman scattering arises from magnetic anisotropies (that gap the magnon spectra at the zone center) and generally gives rise to sharp low-energy Raman features [44–46] in long-range magnetically ordered systems like CrAs. In contrast, two-magnon Raman scattering generally results in broader features [44,47,48]. The large FWHMs of the excitations lead us to suspect their two-magnon origin. As the two-magnon Raman scattering process must fulfill the momentum conservation law and result in a q≈ 0 excitation, it must correspond to the excitation of two magnons with opposite momenta. Those primarily involve regions of the reciprocal space with large magnon density of states (M-DOS), i.e. zone boundaries [49], corresponding to local spin flips in a real-space picture.

In the past, measurements of the two-magnon Raman spectra correctly estimated the exchange coupling constant of the parent compound of high-$T_c$ cuprates [47], which are $S = 1/2$ antiferromagnets on square lattices.

The more general case of a non-collinear magnet presented here is unfortunately more complicated to interpret. The two-magnon energy can in principle be evaluated in two ways. We can either calculate the energy cost of two neighboring spin flips or compute the M-DOS. However, in both cases, difficulties arise from the non-collinearity and the itinerant character of the system, when *e.g.* considering final-state interactions [50]. Additionally, neither the actual spin contribution to the total magnetic moment (1.73 μ$_B$/Cr [8]) nor the exchange coupling constants are precisely known. Regarding the spin value, the large (~60 %) orbital contribution to the total magnetic moment of metallic Cr [51] strongly suggest that the spin in CrAs might be larger than 1/2.

The magnetic structure of CrAs has been previously [7,17] modelled using a Heisenberg-like Hamiltonian $\mathcal{H} = \sum_{i,j} J_{ij} \vec{S}_i \cdot \vec{S}_j$ considering the predominant exchange couplings, which are denoted as $J_a, J_b, J_{c1}$ and $J_{c2}$ following [17] (shown in Fig. 7(c)). The stability conditions for the double helical magnetic ground state have been determined as: $J_{c2}/J_a = 7.1$ and $J_{c1}/J_a = -0.52$ [7], irrespective of the value and sign of $J_b$ [52]. However, the values of the exchange couplings have only been estimated from inelastic neutron scattering data on polycrystalline samples [17], assuming that the total magnetic excitation bandwidth of ~110 meV is dominated by $J_{c2} \gg J_{c1}, J_a$ and amounts to $\sim 2J_{c2}S$.

Based on these estimated exchange parameters, we have calculated the linear spin wave spectra [53] of CrAs and the M-DOS, as shown in Fig. 7(d) in units of $J_{c2}S$. The latter is clearly dominated by the highest energy excitations around $\sim 1.7 J_{c2}S$, which amounts to 93.5 meV (754 cm$^{-1}$) according to the estimate of ref. [17]. Considering two-magnon processes and doubling this value (187 meV or 1508 cm$^{-1}$), we obtain an energy relatively close to that of the feature at 1700 cm$^{-1}$ ≈ 211 meV. In this case, however, none of the other features of this M-DOS can be easily associated with the 350 cm$^{-1}$ mode.

This highlights the fact that linear spin-wave theory is not appropriate to discuss the magnetic excitation spectra of non-collinear magnets. As pointed out in ref. [54], nontrivial anharmonic corrections to the spin-wave spectrum for a non-collinear magnet appear already to the first order in $1/S$ and strongly modify the spin wave dispersion. This has for instance been recently verified experimentally in non-collinear antiferromagnetic CaCr$_2$O$_4$ [55], in which the magnetic excitations are found to be much softer than the predictions from the linear spin-wave theory. Thus these call for experimental investigation of the dispersion of the magnetic excitations in CrAs.

We note that though the magnetic Raman response of Cd$_2$Os$_2$O$_7$ shows a good agreement with its M-DOS [56], difficulties have been encountered in interpreting the two-magnon Raman data of other non-collinear magnets such as α-SrCr$_2$O$_4$ [57] or Bi$_2$Fe$_4$O$_9$ [58]. To the best of our knowledge, the impact of these effects on the magnetic Raman response have only been evaluated for the case of a triangular lattice [59,60]. However, they are strongly model dependent and should be evaluated for the specific magnetic structure of CrAs.

The present experimental results will provide a strong benchmark for future experimental and theoretical investigations.

## VI. Summary and Conclusion

In summary, we have reported a detailed Raman scattering study of lattice and magnetic excitations as functions of temperature in CrAs single crystals. We detected abrupt changes in phonon frequency for all A$_g$ modes across $T_N$ =265 K, at which CrAs undergoes a first-order magneto-structural phase transition characterized by the formation of a double helical magnetic structure, which is accompanied by a large expansion of the crystal unit cell. We further reported evidences of magnetic light scattering in this itinerant non-collinear magnet.

First principle lattice structure and dynamics calculations (constrained to the experimental lattice parameters) show a very good agreement with the data, in particular when magnetic order is considered. This indicates a sizeable spin-phonon coupling in CrAs, but further work is required to really quantify these effects as we are, in particular, limited to collinear magnetic structures.

Furthermore, calculations beyond the harmonic level are needed to understand the temperature dependence of the linewidth of the strongest A$_g$ mode which appears highly anomalous. It shows an unusual broadening in the magnetic phase down to 125 K, which might arise from its coupling to magnetic degrees of freedoms. To discuss the potential relevance of these effects to superconductivity, it would be insightful to track the phonon anomalies and the magnetic excitations as a function of hydrostatic pressure or doping (e.g. CrAs$_{1-x}$P$_x$) as the magnetic order gets suppressed.

**Acknowledgements**
We acknowledge C. Bernhard, R. Eder, M. Garst, C. Meingast, I. Paul, J. Schmalian, Q. Si and S-M. Souliou for valuable discussions. The contribution from M.M. was supported by the Karlsruhe Nano Micro Facility (KNMF). K.W. acknowledges funding from the Alexander von Humboldt Foundation.

# APPENDIX A: EXPERIMENTAL STRUCTURAL PARAMETERS AT 295 AND 95 K

Table I summarizes the structural parameters of CrAs single crystals at 295 and 95 K. The detail procedure to obtain such parameters is mentioned in Sec. II A.

Table I. Structural parameters of CrAs determined from temperature-dependent single-crystal x-ray diffraction. The structure was refined in the orthorhombic space group *Pnma*. Cr as well as As reside on Wyckoff positions $4c$ with coordinates $x$, ¼, $z$. The $U_{ii}$ denote the atomic displacement

|    |                     | 295 (K)    | $T = 95$ (K) |
|----|---------------------|------------|--------------|
|    | $a$ (Å)             | 5.6535(5)  | 5.6095(6)    |
|    | $b$ (Å)             | 3.4737(4)  | 3.5905(5)    |
|    | $c$ (Å)             | 6.2042(6)  | 6.1384(6)    |
| Cr | $x$                 | 0.0064 (2) | 0.0073(2)    |
|    | $z$                 | 0.2012(1)  | 0.2056(1)    |
|    | $U_{11}$ (Å$^2$)    | 0.0048(5)  | 0.0024(4)    |
|    | $U_{22}$ (Å$^2$)    | 0.0107(5)  | 0.0052(5)    |
|    | $U_{33}$ (Å$^2$)    | 0.0064(5)  | 0.0044(4)    |
|    | $U_{13}$ (Å$^2$)    | 0.0001(3)  | -0.0004(3)   |
|    | $U_{eq}$ (Å$^2$)    | 0.0073(3)  | 0.0040(2)    |
|    | SOF                 | 0.9988(55) | 0.9984(52)   |
| As | $x$                 | 0.2022(1)  | 0.2054(1)    |
|    | $z$                 | 0.5770(1)  | 0.5836(1)    |
|    | $U_{11}$ (Å$^2$)    | 0.0075(3)  | 0.0048(3)    |
|    | $U_{22}$ (Å$^2$)    | 0.0076(3)  | 0.0026(3)    |
|    | $U_{33}$ (Å$^2$)    | 0.0075(3)  | 0.0028(3)    |
|    | $U_{13}$ (Å$^2$)    | 0.0010(3)  | 0.0004(2)    |
|    | $U_{eq}$ (Å$^2$)    | 0.0075(2)  | 0.0034(2)    |
|    | $wR_2$              | 4.15       | 3.98         |
|    | $R_1$               | 1.90       | 1.66         |

factors ($U_{12}$[Cr]= $U_{23}$[Cr]= $U_{12}$[As]= $U_{23}$[As]= 0). Refinement of the site occupancy factor (SOF) of Cr shows that the sample is stoichiometric within the error bars (The SOF of As was fixed to 1).

# APPENDIX B: RAMAN SPECTRA IN CROSSED POLARIZATION

The high-resolution Raman spectra in crossed-polarization configuration are extremely weak (not shown), in which the incoming photon polarization is parallel to the crystallographic *a*-axis of the crystal. The intensity (I) of the strongest mode in this configuration amounts to I=0.2 counts/sec at 225 K; whereas at the same temperature, the strongest A$_g$ phonon in parallel-polarization has I=5 counts/sec.

However, we could resolve at least four phonon modes of B$_{1g}$ and B$_{2g}$ symmetries in the Raman spectra measured with the low-resolution grating at 10 and 310 K, as shown in Fig. 8. We didn't observe any magnetic Raman scattering in this configuration.

# APPENDIX C: RAMAN SPECTRA FROM THE SIX FACETS OF A CrAs CRYSTAL

Fig. 9 shows the Raman spectra in XX geometry, which were obtained from the six facets of a physical CrAs crystal. In a good agreement with the relevant polarization selection rules, we always observed only A$_g$ modes irrespective of the direction of the incoming photons.

# APPENDIX D: Analysis of Fleury-Loudon Hamiltonian

The effective two-magnon light scattering operator can be written as

$$\hat{O} = \sum_{i,j} \eta_{ij} (\vec{E}_i \cdot \vec{d}_{ij})(\vec{E}_s \cdot \vec{d}_{ij}) \vec{S}_i \cdot \vec{S}_j,$$

where, $\vec{E}_i$ ($\vec{E}_s$) is the polarization of the incident (scattered) light, $\vec{d}_{ij}$ is the unit vector connecting the magnetic sites $i$ and $j$, on which the spins $\vec{S}_i$ and $\vec{S}_j$ respectively sit, and $\eta_{ij}$ is a matrix element proportional to exchange coupling constant [61]. The sum runs over all pairs of magnetic atoms. The relevant magnetic sites are marked in Fig. 7(c). The angle between $\vec{E}_{in}$ (parallel to *a*-axis) and Cr1-Cr2 or Cr1-C3 is very close to 90º. Thus we can neglect the projections of $\vec{d}_{12}$ and $\vec{d}_{13}$ vectors on the crystallographic *a*-axis. In the chosen scattering geometry, the main contribution arises from scattering across the Cr1-Cr4 bond ($\vec{d}_{14}$).

For the XX-geometry, we get:
$$\hat{O}_{XX} \sim \eta_{14} \cos^2(\theta_{14}) \vec{S}_1 \cdot \vec{S}_4,$$
where, $\theta_{14}(\approx 11º)$ is the angle between *a*-axis and $\vec{d}_{14}$. For the XY-geometry, we obtain:
$$\hat{O}_{XY} \sim \eta_{14} \cos(\theta_{14}) \sin(\theta_{14}) \cos^2(\alpha) \vec{S}_1 \cdot \vec{S}_4,$$
in which $\alpha \approx 30º$, since the orientation of the measured facet is (011).

The ratio between the magnetic scattering intensities of the two channels will therefore be
$$\frac{I_{XY}}{I_{XX}} \sim |\tan(\theta_{14}) \cos^2(\alpha)|^2 \sim 2 \times 10^{-2}.$$

Such a ratio is consistent with the absence of magnetic scattering signal in the XY-geometry.

* kaushik.sen@kit.edu
† matthieu.letacon@kit.edu

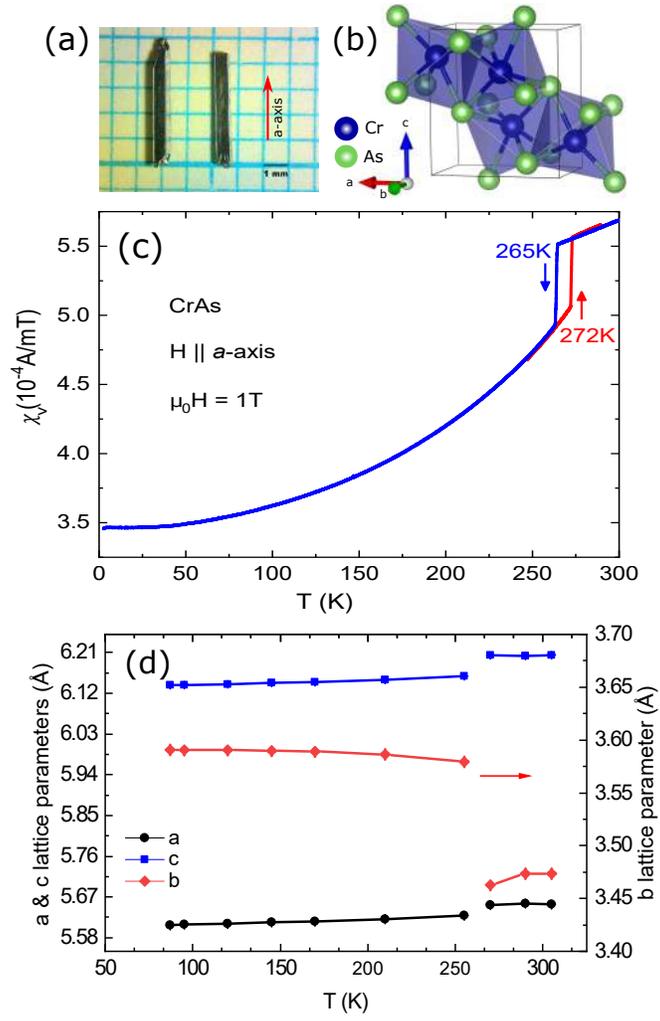

FIG. 1. (a) Typical size and morphology of as-grown single crystals of CrAs. The crystallographic *a*-axis is marked with respect to the crystals. (b) Crystallographic unit cell of orthorhombic CrAs (space group: *Pnma*, no. 62). (c) Magnetic volume susceptibility ($\chi_v$) of CrAs as a function of temperature during cooling and warming. The large drop in $\chi_v$ at 265 K during cooling marks the first-order magneto-structural transition. (d) *a*-, *b*- and *c*-axis lattice parameters as functions of temperature, which were determined from single-crystal x-ray diffraction.

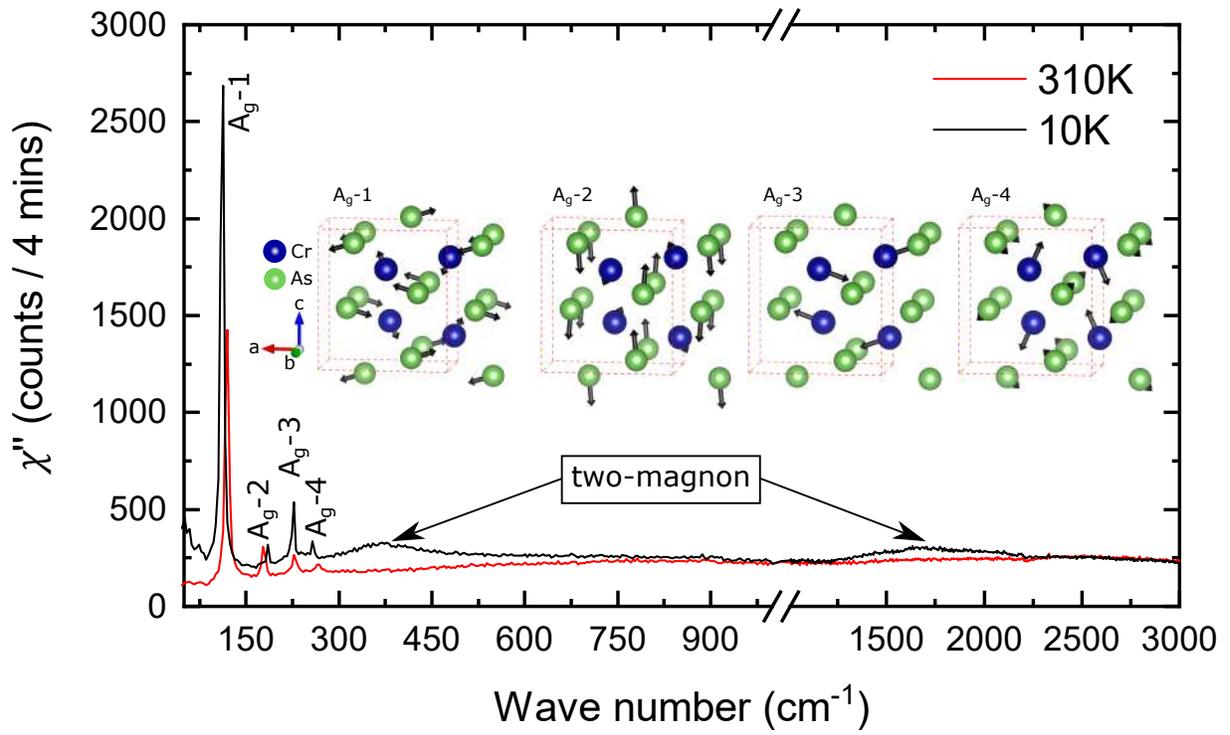

FIG. 2. Representative Raman spectra ($\chi''$) at two temperatures below (10 K) and above (310 K) $T_N = 265$ K in XX geometry which probes the lattice dynamics and magnetic excitations of $A_g$ symmetry. The sharp peaks at low energy are four phonon modes of $A_g$ symmetry. The two broad modes at higher energy originate from magnetic Raman scattering which we attribute to two-magnon modes. Insets show the atomic displacement patterns of the four phonon modes, which were determined from the DFT-based lattice dynamics calculations.

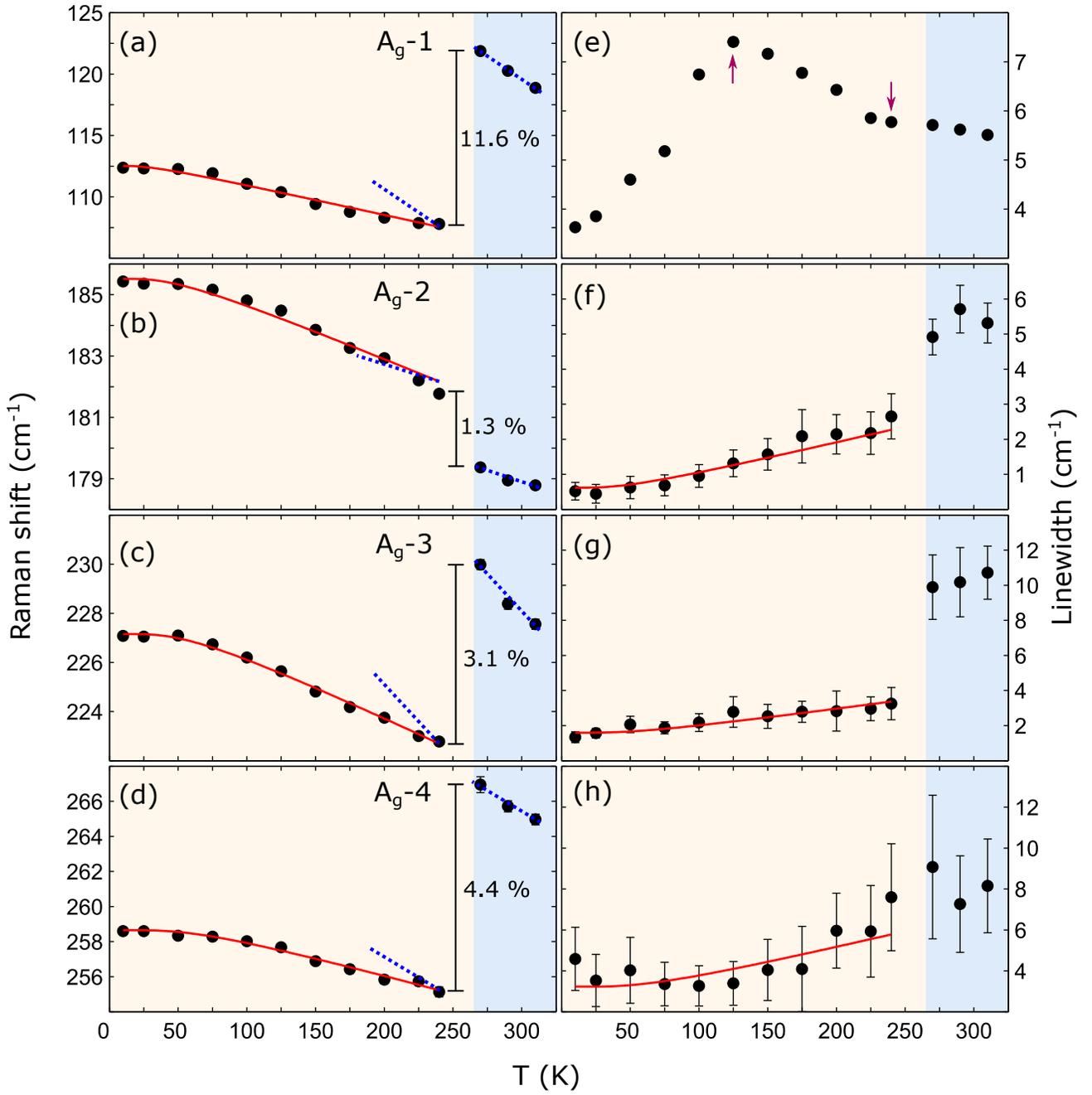

FIG. 3. (a)-(d) Raman shift as a function of temperature for the four $A_g$ phonon modes. (e)-(h) The corresponding linewidths as functions of temperature. For small uncertainty the errorbars are hidden behind the symbols. The solid lines (red) are the best fits to the data with the Klemens decay model [30, 31]. The dotted lines in (a)-(d) in low-temperature phase ($< T_N = 265$ K) are parallel to the corresponding lines which go through the data in the high-temperature phase. The down-arrow in Fig. 3(e) marks the onset temperature near $T_N$, below which the linewidth starts broaden upon cooling down to 125 K (indicated by the up-arrow).

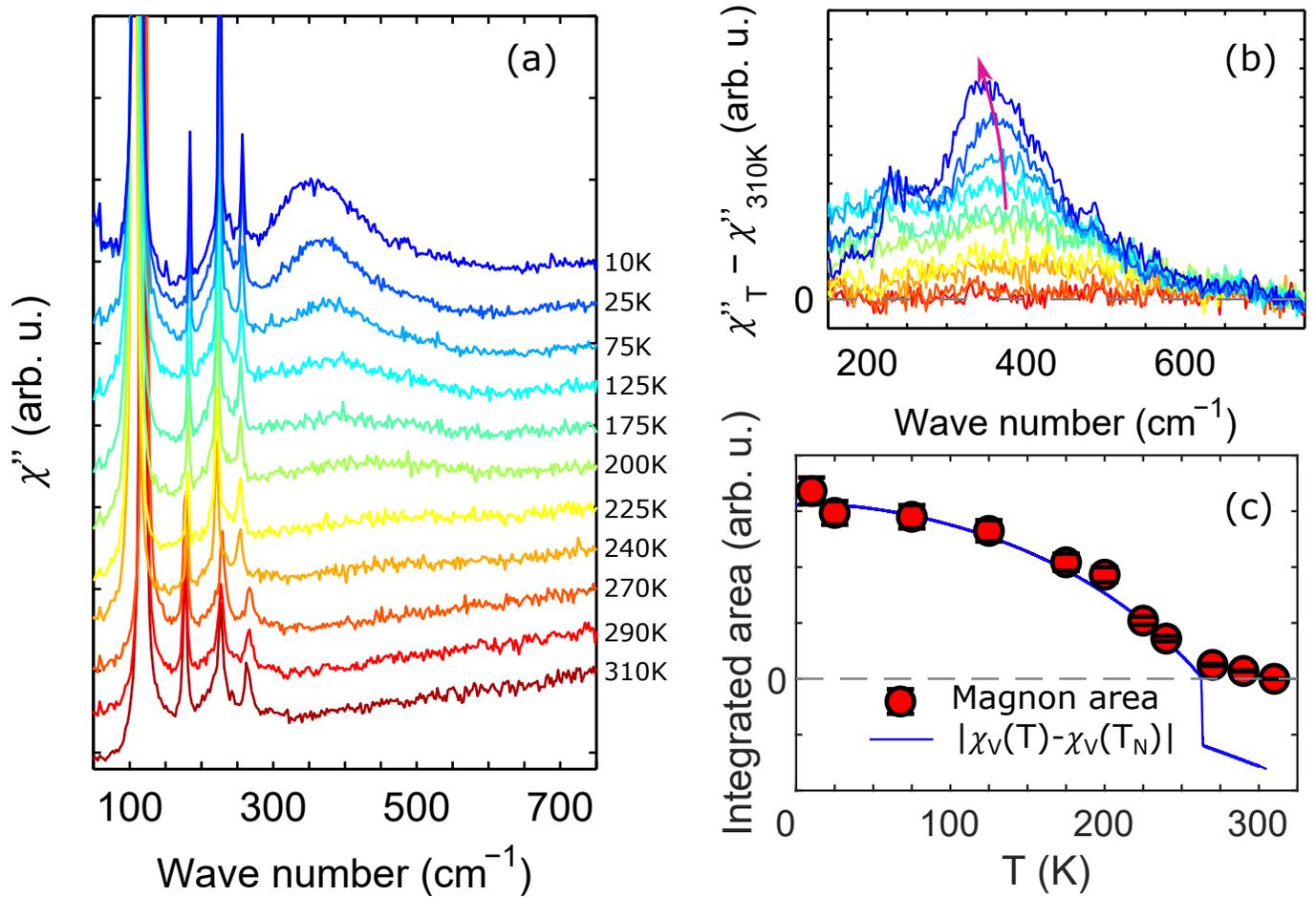

FIG. 4. (a) Normalized Raman response ($\chi''$) as a function of temperature. (b) $\Delta\chi''(T) = \chi''(T) - \chi''(310K)$ at several temperatures. Prior to the calculation of $\Delta\chi''(T)$, we first subtracted the phonon contributions from the individual spectrum. The arrow marks the apparent softening of the most intense mode. (c) Integrated area under the curves in (b) as a function of temperature. The solid line through the data is evaluated from $\chi_v$ (cooling curve), as described in the text.

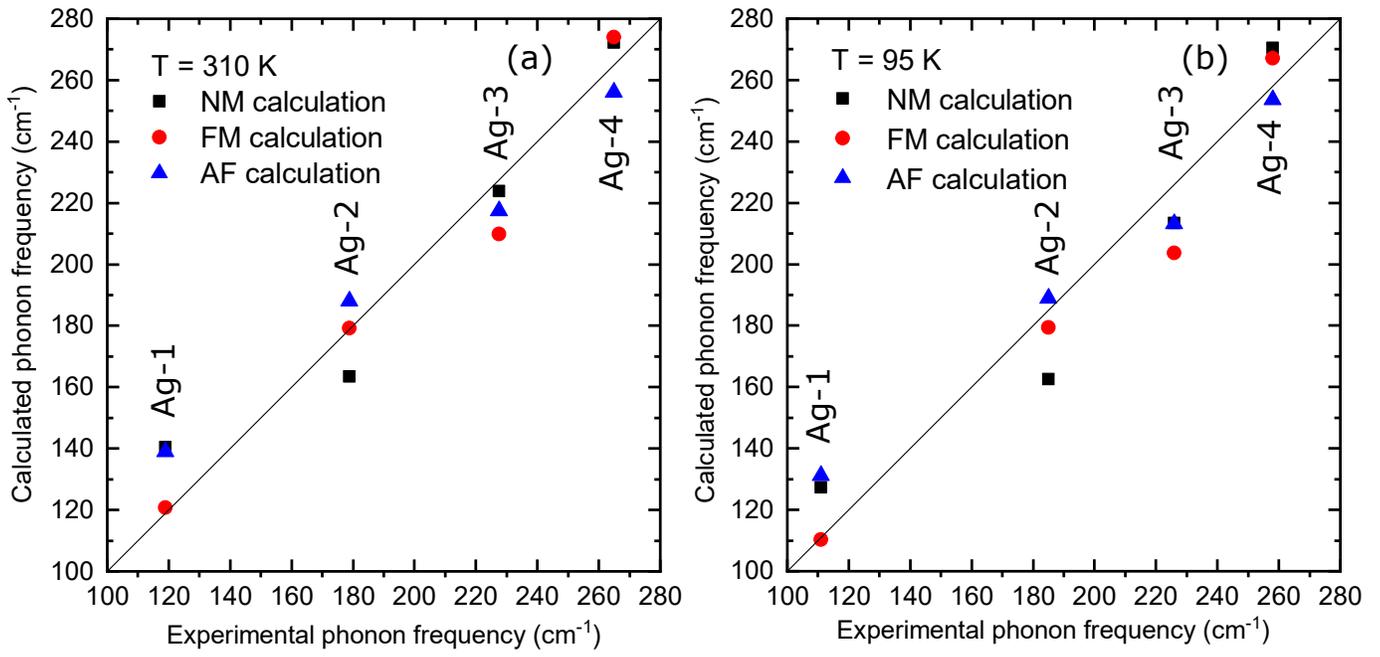

FIG. 5. Calculated phonon frequencies in non-magnetic (NM), ferromagnetic (FM) and antiferromagnetic (AF) environments are plotted against the experimental phonon frequencies at (a) 310 K and (b) 95 K. The straight lines through the data have the slopes= 1, which means data points fall on these lines if the experimental and calculated phonon frequencies are equal.

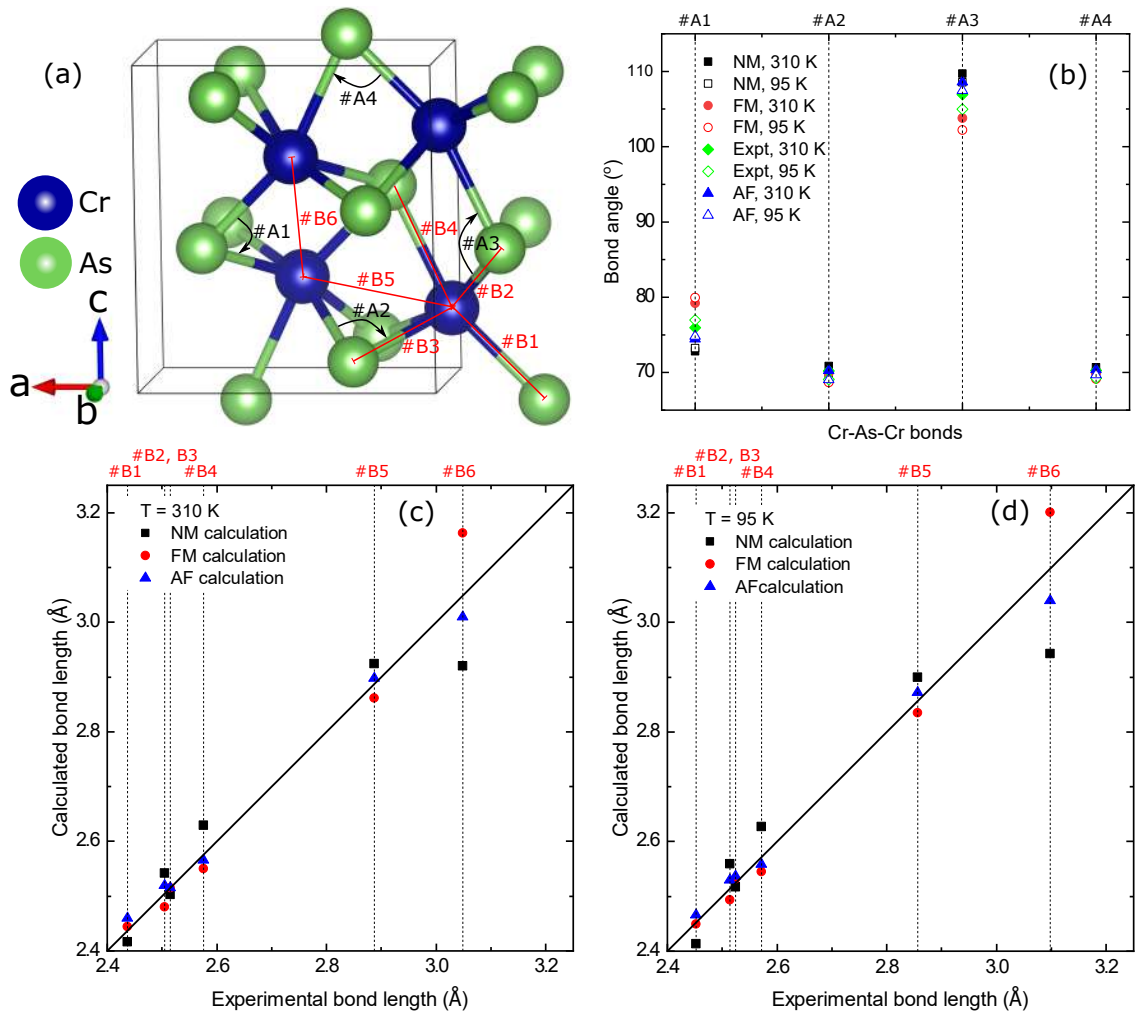

FIG. 6. (a) A sketch labeling the interatomic distances and Cr-As-Cr bond angles. (b) Comparison of the experimental bond angles with the calculations at 310 and 95 K in non-magnetic (NM), ferromagnetic (FM) and antiferromagnetic (AF) cases. (c) Comparison of the interatomic distances at 310 K with the calculations. (d) The corresponding plot at 95 K.

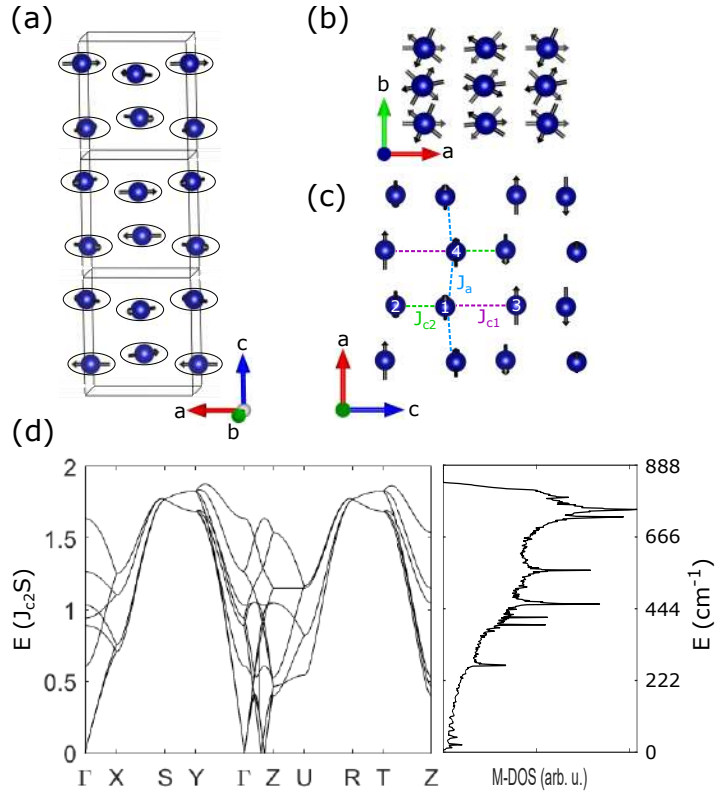

FIG. 7. (a) The double helical magnetic ground state of CrAs with the propagation vector of $Q_z = 0.36 \times 2\pi c^*$. (b) A view of the magnetic structure along $c$-axis. (c) Magnetic exchange couplings that are required to stabilize such a magnetic ground state are marked on the crystallographic $ac$-plane of CrAs. (d) Left: Magnon dispersion along high-symmetry directions in reciprocal space. Right: Magnon density of states (M-DOS). $J_{c2}$ and $S$ in energy units are the strongest exchange coupling and the spin state of Cr, respectively. According to the estimation of $2J_{c2}S \approx 110$ meV from inelastic neutron scattering data [17], an additional energy scale is given in cm$^{-1}$ for M-DOS.

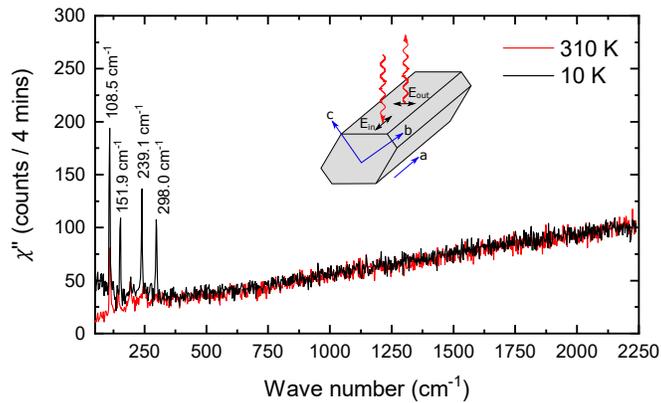

FIG. 8. Low-resolution Raman spectra in crossed polarization geometry at 10 and 310 K. The incoming photon polarization was parallel to the crystallographic $a$-axis. This polarization geometry probes phonon modes of $B_{1g}$ and $B_{2g}$ symmetries. No magnetic scattering is visible in this polarization-geometry. Inset schematically shows the polarization geometry with respect to the corresponding crystal facet, in which incoming ($E_{in}$) and outgoing ($E_{out}$) polarizations are marked.

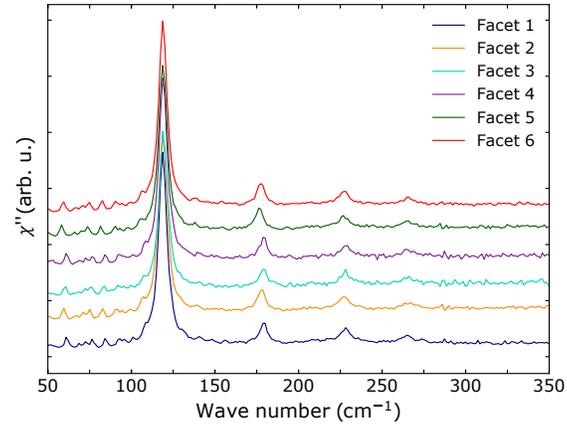

FIG. 9. Raman spectra in XX geometry at room temperature from the six facets of a CrAs crystal.